\documentclass[preprint,amsmath,amssymb,nofootinbib]{revtex4}

\usepackage{graphicx}
\usepackage{dcolumn}
\usepackage{bm}

\begin{document}

\title{Obtaining  a closed-form representation for the dual bosonic thermal Green function by using methods of integration on the complex plane}

\author{Leonardo Mondaini}
\email{mondaini@unirio.br}
\affiliation{Grupo de F\'{i}sica Te\'{o}rica e Experimental, Departamento de Ci\^encias Naturais, Universidade Federal do Estado do Rio de 
Janeiro, Av. Pasteur 458, Urca, Rio de Janeiro-RJ 22290-240, Brazil}


\begin{abstract}
We derive an exact closed-form representation for the Euclidean
thermal Green function of the two-dimensional (2D) free massless
scalar field in coordinate space. This can be interpreted as the
real part of a complex analytic function of a variable
that conformally maps the infinite strip $-\infty<x<\infty$
($0<\tau<\beta$) of the $z=x+i\tau$ ($\tau$: imaginary time) plane
into the upper-half-plane. Use of the Cauchy-Riemann conditions, then
allows us to identify the dual thermal Green function as the
imaginary part of that function.

{\bf{Keywords:}} thermal Green function, massless scalar field, residue theorem

\end{abstract}


\maketitle


\section{Introduction}
As remarked in \cite{weldon1}, despite the fact that quantum field
theories are usually formulated in coordinate space, calculations,
in both $T=0$ and $T\neq 0$ cases, are almost always performed in
momentum space. However, when we are faced with the exact
calculation of correlation functions we are naturally led to the
problem of finding closed-form expressions for Green functions in
coordinate space \cite{delepine, MondainiMarinoJPA}.

A closed-form representation for the thermal
Green function of the free massless scalar field\footnote{A field $\phi(x)$ is called a \emph{scalar} field, in contrast to a \emph{vector} or \emph{tensor} field,  when under a Lorentz transformation $x^\mu \rightarrow x'^\mu = \Lambda^\mu \,_\nu x^\nu$, it transforms, trivially, according to $\phi(x) \rightarrow \phi'(x) = \phi(\Lambda^{-1}x)$. The quantization of such fields gives rise to spin-0 particles (\emph{scalar bosons}), like the famous Higgs boson (a proposed elementary particle in the Standard Model of particle physics). However, it is a well-known fact that most particles in the universe have a nonzero intrinsic spin. These particles arise in field theory when we consider fields which transform non-trivially under Lorentz transformations. Indeed, fields with spin have more complicated transformation laws, since the various components of the fields rotate into one another under Lorentz transformations. A good example of such a field is the vector field $A^\mu(x)$ of electromagnetism, which transforms as
$A^\mu(x) \rightarrow \Lambda^\mu \,_\nu A^\nu(\Lambda^{-1}x)$. In field theory, spin-1 particles are described as the quanta of vector fields. Such \emph{vector bosons} play a central role as the mediators of interactions (\emph{force carriers}) in particle physics. Another good example comes from the Dirac equation, whose quantization gives rise to spin-1/2 particles (\emph{fermions}). \\  Last but not least, we must stress that, for each kind of field, we may conceive field theories describing massive and massless particles. For instance, in the case of vector (spin-1) fields, we may cite as important examples the gauge fields of the electromagnetic, the weak, and the strong interactions, whose corresponding vector bosons are, respectively, massless photons, massive $W^\pm$ and $Z^0$ bosons, and massless gluons.} in 2D was firstly presented in \cite{delepine}, where attention has been focused on its
relation with the bosonization of the massive Thirring model (MTM)\footnote{The MTM is described by the Lagrangian density
${\mathcal L} = \bar\psi(i\gamma_\mu\partial^\mu - m)\psi - \frac{g}{2}(\bar \psi \gamma_\mu \psi)(\bar \psi \gamma^\mu\psi)$, where $\psi$ is a two-component Dirac fermion field in
(1+1)-dimensions, and  $\gamma^\mu$ are Dirac gamma matrices. Notice that the interaction is the only local interaction possible since the model involves only four fermion variables. It is well known that it can be mapped, by a method called bosonization, into the sine-Gordon model of a scalar field, whose dynamics is determined by
${\mathcal L}  = \frac{1}{2}\partial_\mu\phi\
\partial^\mu\phi +2\alpha \cos\eta\phi$, where the couplings in the two models are related as $g = \pi \left (4\pi/\eta^2 - 1 \right)$; $m \bar \psi \psi = - 2\alpha \cos\eta\phi$ .  Both models have been extensively studied.} using the imaginary-time formalism for finite temperature quantum field theory \cite{Das}. Unfortunately,  the authors omit valuable details of the calculations, which would be useful for graduate students and researchers working on this subject.

In the present work we present a simple and yet
appealing step-by-step derivation of an exact closed-form representation for the
thermal Green function of 2D free massless scalar theory in the
coordinate space, at a level accessible to usual graduate students in physics. This has been obtained by using the imaginary-time formalism along with methods of integration on the complex plane and the software \emph{Mathematica}. The peculiar form of this, allows us to easily
recognize it as the real part of an analytic function, a
fact that leads us to determine the corresponding dual thermal Green
function as the imaginary part of that function, according to the
Cauchy-Riemann conditions. This dual thermal Green function turns
out to be a key ingredient for the obtainment of
fermion correlators of the MTM at finite temperature, as shown in \cite{MondainiMarinoMPLA}.

\section{Two-dimensional thermal Green function}

The Euclidean thermal Green function of the 2D free
massless scalar theory can be written in the coordinate space (${\bf{r}}\equiv
(x,\tau)$) as \cite{Das}
\begin{equation}
G_T({\bf{r}})=\frac{1}{\beta}\sum_{n=-\infty}^{\infty}\int_{-\infty}^{\infty}
\frac{dk}{2\pi}\frac{e^{-i(kx+\omega_n\tau)}}{k^2+\omega_n^2},
\label{gft1}
\end{equation}
where $\omega_n=2\pi n/\beta$, $\beta=1/k_BT$. At $T=0$, $G_T({\bf{r}})$ reduces to the
2D Coulomb potential.

\begin{figure} [h!]
\hfil\scalebox{0.4}{\includegraphics{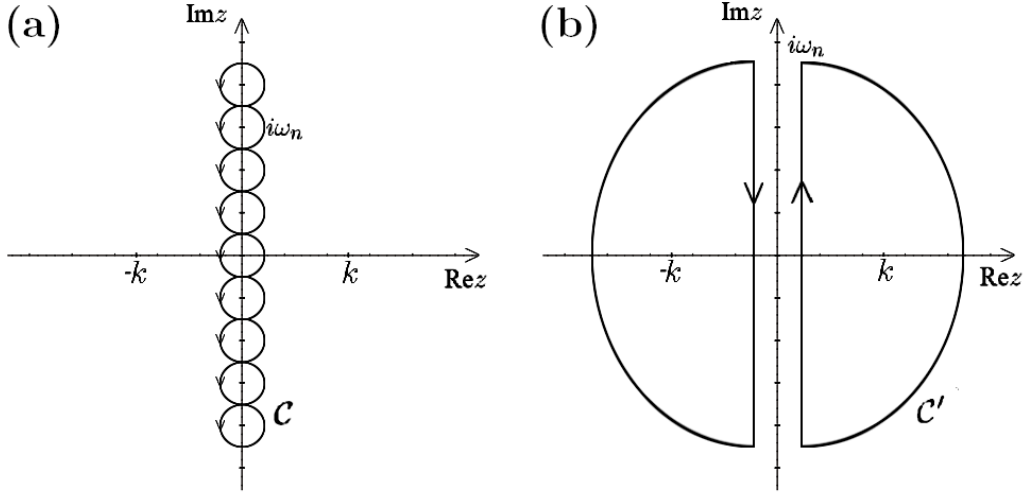}}\hfil
\caption{Integration contour in the complex plane used in (a) Eq.
(\ref{gft5}) and (b) Eq. (\ref{gft6}).} \label{contour}
\end{figure}

The sum appearing in (\ref{gft1}) may be evaluated by considering
the following integral on the complex plane \cite{FW}
\begin{equation}
I_\mathcal{C}=\frac{1}{2\pi i}\oint_\mathcal{C}
dz\,f(z)\delta_{BE}(\beta z), \label{gft3}
\end{equation}
where
\begin{equation}
f(z)=\frac{e^{z\tau}}{k^2-z^2}, \ \ \ \ \ \ \ \ \ \ \ \
\delta_{BE}(\beta z)=\frac{1}{e^{\beta z}-1}, \label{gft4}
\end{equation}
and the integration contour $\mathcal{C}$ is defined in Fig.
\ref{contour}-(a).

The function $f(z)$ has poles at $z=\pm k$, being therefore outside
the contour $\mathcal{C}$. Those of $\delta_{BE}(\beta z)$, are
situated at $z=i\,2\pi n/\beta=i\omega_n$, ($n=0, \pm 1, \pm 2,
\ldots$) hence inside the contour $\mathcal{C}$.



It is easy to see, using the residue theorem \cite{Arfken} that the sum coincides
with the integral $I_\mathcal{C}$ and, therefore, we have

\begin{equation}
\begin{split}
I_\mathcal{C}&=\sum_{n=-\infty}^{\infty}\textrm{Res}\left[f(i\omega_n)\delta_{BE}(\beta
(i\omega_n))\right] \\
&=\frac{1}{\beta}\sum_{n=-\infty}^{\infty}\frac{e^{-i\omega_n\tau}}{k^2+\omega_n^2}.
\end{split}
\label{gft5}
\end{equation}

Deforming the contour  $\mathcal{C}$ into $\mathcal{C'}$ shown in
Fig. \ref{contour}-(b) we have, using the residue theorem again:
\begin{equation}
\begin{split}
I_\mathcal{C'}&=-\left(\textrm{Res}\left[f(-k)\delta_{BE}(\beta
(-k))\right]+\textrm{Res}\left[f(k)\delta_{BE}(\beta
k)\right]\right)\\
&=\frac{\cosh
\left[k\left(\tau-\frac{\beta}{2}\right)\right]}{2k\,\sinh
\left[k\left(\frac{\beta}{2}\right)\right]}. \label{gft6}
\end{split}
\end{equation}

Hence, since $I_\mathcal{C}=I_\mathcal{C'}$, we have
\begin{equation}
\frac{1}{\beta}\sum_{n=-\infty}^{\infty}\frac{e^{-i\omega_n\tau}}{k^2+\omega_n^2}=
\frac{\cosh
\left[k\left(\tau-\frac{\beta}{2}\right)\right]}{2k\,\sinh
\left[k\left(\frac{\beta}{2}\right)\right]},\label{??}
\end{equation}
which allows us to rewrite the expression (\ref{gft1}) for $G_T({\bf{r}})$ as
\begin{equation}
G_T({\bf{r}})=\frac{1}{4\pi}\int_{-\infty}^{\infty} dk\,\frac{e^{-i
kx}\,\cosh \left[k\left(\tau-\frac{\beta}{2}\right)\right]}{k\,\sinh
\left[k\left(\frac{\beta}{2}\right)\right]}. \label{gft8}
\end{equation}

Observe that only the real part of the integral is non-vanishing.
Making the change of variable $k\rightarrow (2\pi/\beta)y$, defining
$a\equiv (2\pi/\beta)x$,
$\theta\equiv\left[(2\pi/\beta)\tau-\pi\right]$, and introducing the
regulator $b$  we may rewrite (\ref{gft8}) as
\begin{equation}
G_T({\bf{r}})=\lim_{b\rightarrow 0}\frac{1}{2\pi}\int_{0}^{\infty}
dy\,\frac{y \cos ay\,\cosh \theta y}{\left(y^2+b^2\right)\,\sinh \pi
y}. \label{gft16}
\end{equation}

This integral, which can be found in \cite{grad}, gives
\begin{equation}
G_T({\bf{r}})=\lim_{b\rightarrow 0}\left\{\frac{e^{-|a|b}\cos
b\theta}{4\sin b\pi}+\sum_{k=1}^{\infty}(-1)^k \frac{k e^{-|a|k}\cos
k\theta}{2\pi(k^2-b^2)}\right\}. \label{gft17}
\end{equation}

The sum above may be evaluated in terms of hypergeometric functions
$_2F_1$ with the help of {\emph {Mathematica}} \cite{Mathematica}. The result is
\begin{widetext}
\begin{equation}
\begin{split}
G^T({\bf{r}})=\lim_{b\rightarrow 0} &
\left\{\frac{e^{-|a|b}\cos b\theta}{4\,\sin b\pi}+
\frac{e^{-|a|-i \theta}}{8\pi \left(b^2-1\right)} \left[ b \,
_2F_1\left(1-b,1,2-b;-e^{-|a|-i \theta }\right) \right .\right .\\
& \left .\left .+\, _2F_1\left(1-b,1,2-b;-e^{-|a|-i \theta }\right) +b
e^{2 i \theta } \, _2F_1\left(1-b,1,2-b;-e^{-|a|+i \theta}\right)
\right .\right .\\
& \left .\left .+e^{2 i \theta } \, _2F_1\left(1-b,1,2-b;-e^{-|a|+i
\theta}\right) -b \, _2F_1\left(b+1,1,b+2;-e^{-|a|-i \theta }\right)
\right .\right .\\
& \left .\left .+\, _2F_1\left(b+1,1,b+2;-e^{-|a|-i \theta }\right) -b
e^{2 i \theta } \, _2F_1\left(b+1,1,b+2;-e^{-|a|+i \theta}\right)
\right .\right .\\
& \left .\left .+e^{2 i \theta } \, _2F_1\left(b+1,1,b+2;-e^{-|a|+i
\theta}\right)\right]\right\}.
\end{split}
\label{gft18}
\end{equation}
\end{widetext}
Taking the $b\rightarrow 0 $ limit in the expression above, we obtain (since $_2F_1\left(1, 1, 2; -z\right) = \frac{\ln(1+z)}{z}$)
\begin{equation}
G_T({\bf{r}}; b)=\frac{1}{4\pi
b}-\frac{|a|}{4\pi}-\frac{1}{4\pi}\ln\left[\left(1+e^{-|a|-i\theta}\right)\left(1+e^{-|a|+i\theta}\right)\right].
\label{gft19}
\end{equation}

By inserting the expressions for $a$ and $\theta$ and
defining the regulator mass $\mu_0\equiv (2\pi/\beta)e^{-1/2b}$, we
may write the scalar thermal Green function, after some algebra, as
\begin{widetext}
\begin{equation}
G_T({\bf{r}})=\lim_{\mu_0\rightarrow
0}-\frac{1}{4\pi}\ln\left\{\frac{\mu_0^2\beta^2}{\pi^2}\left[\cosh\left(\frac{2\pi}{\beta}x\right)-\cos\left(\frac{2\pi}{\beta}\tau\right)\right]\right\},
\label{gft20nova}
\end{equation}
\end{widetext}
which coincides with the result presented for the first time in \cite{delepine}.

Finally, notice that the Eq. (\ref{gft20nova}) may be also written as
\begin{equation}
G_T({\bf{r}})=\lim_{\mu_0\rightarrow
0}-\frac{1}{4\pi}\ln\left[\mu_0^2\zeta({\bf{r}})\zeta^*({\bf{r}})\right], \label{gft20}
\end{equation}
where $(z=x+i\tau)$
\begin{equation}
\zeta({\bf{r}})\equiv
\zeta(z)=\frac{\beta}{\pi}\sinh\left(\frac{\pi}{\beta}z\right).
\label{zeta}
\end{equation}

\section{The dual thermal Green function}

From Eq. (\ref{gft20}) we can also see that the thermal Green
function may be written as the real part of an analytic function of a complex
variable $\zeta$, namely
\begin{equation}
G_T({\bf{r}};\mu_0)=\textrm{Re}\left[\mathcal{F}(\zeta)\right]=\frac{1}{2}\left[\mathcal{F}(\zeta)+\mathcal{F}^*(\zeta)\right],
\label{gft20b}
\end{equation}
where $\mathcal{F}(\zeta)\equiv -(1/2\pi)\ln\left[\mu_0\zeta({\bf{r}})\right]$.

The imaginary part of $\mathcal{F}(\zeta)$ may be written as
\begin{equation}
\begin{split}
\tilde{G}_T({\bf{r}})\equiv\textrm{Im}\left[\mathcal{F}(\zeta)\right]&=\frac{1}{2i}
\left[\mathcal{F}(\zeta)-\mathcal{F}^*(\zeta)\right]\\&=-\frac{1}{4\pi
i}\ln\left[\frac{\zeta({\bf{r}})}{\zeta^*({\bf{r}})}\right].
\label{gft20c}
\end{split}
\end{equation}

Now, from the analyticity of $\mathcal{F}(\zeta)$, then, it follows
that its imaginary and real parts must satisfy the Cauchy-Riemann conditions, which are given by
\begin{equation}
\epsilon^{\mu\nu}\partial_\nu G_T =-\partial_\mu \tilde{G}_T, \ \ \
\ \ \ \ \ \ \ \ \ \epsilon^{\mu\nu}\partial_\nu
\tilde{G}_T=\partial_\mu G_T. \label{cauchyriemann}
\end{equation}
This property characterizes $\tilde{G}_T$ as the dual thermal Green
function.

\section{Concluding remarks}
We would like to make a few comments about (\ref{gft20}).
Firstly, we note that in the zero temperature limit ($T\rightarrow
0$, $\beta\rightarrow \infty$), we have $\zeta (z)\rightarrow z $
and $\zeta^*(z)\rightarrow z^* $ and, therefore, we recover the well-known Green function at zero
temperature, namely
\begin{equation}
\lim_{\beta\rightarrow\infty}G_T({\bf{r}};\mu_0)=-\frac{1}{4\pi} \ln
\left[\mu_0^2\, zz^*\right]=-\frac{1}{4\pi} \ln \left[\mu_0^2 ||{\bf{r}}||^2 \right]. \label{gft22}
\end{equation}

Comparing (\ref{gft20}) with (\ref{gft22}) we can also see that the only
effect of introducing a finite temperature is to exchange the complex variable
$z$ for $\zeta(z)$. Since $\zeta(z)$ is analytic, we
conclude that the thermal Green function is obtained from the one at
zero temperature by the following conformal mapping \cite{Arfken}:
the infinite strip $0<\tau<\beta$ and $-\infty<x<\infty$ is mapped into the
region within the upper-half-$\zeta$-plane. Notice that only the values
$[0,\beta]$ of $\tau$ are relevant because this variable is periodic
in $\beta$, as it should at finite $T$.

\section*{Acknowledgments}

The author would like to thank E. C. Marino for his valuable comments during the calculations.

This work has been supported in part by Funda\c c\~ao CECIERJ.


%
%
\end{document}